\documentclass[aps,prd,floats,floatfix,nofootinbib,twocolumn]{revtex4-1}
\usepackage{graphicx}

\begin{document}

\title{Probe of new light Higgs bosons from bottomonium $\chi_{b0}$ decay}

\author{Stephen Godfrey}
\email{godfrey@physics.carleton.ca}

\author{Heather E.~Logan}
\email{logan@physics.carleton.ca}

\affiliation{Ottawa-Carleton Institute for Physics, Carleton University, 1125 Colonel By Drive, Ottawa, Ontario K1S 5B6, Canada}

\date{February 3, 2016}                                  % Activate to display a given date or no date

\begin{abstract}
We calculate the branching ratios of $\chi_{b0}\to \tau^+\tau^-$ via an $s$-channel Higgs boson and estimate the sensitivity to this process from $\Upsilon \to \gamma \chi_{b0} \to \gamma \tau^+\tau^-$.  We show that future running at the $\Upsilon(3S)$ at a very high luminosity super $B$ factory can put significant constraints on the Type-II two-Higgs-doublet model when the discovered 125~GeV Higgs boson is the heavier of the two CP-even scalars.
\end{abstract}

\maketitle 

%%%%%%%%%%%%%%%%%%%%%%%%%%%%%%%%%%%%%%%%%%%%%%
\section{Introduction}

The search for evidence for new physics beyond the Standard Model (SM) has been intensifying
with the discovery of a scalar boson whose properties are consistent
with those expected for the SM Higgs boson.  
Many extensions of the SM include an extended Higgs sector and there has been considerable experimental effort searching for evidence of this.  One possibility to constrain such extensions is through the effects of off-shell Higgs bosons in bottom meson decays to third-generation leptons.  The decay of $B^+ \to \tau^+ \nu$~\cite{Hou:1992sy} (or $B \to \tau \nu X$~\cite{Krawczyk:1987zj,Kalinowski:1990ba,Grossman:1994ax}) is sensitive to $s$-channel exchange of charged Higgs bosons, and $\eta_b \to \tau^+\tau^-$~\cite{Rashed:2010jp} is sensitive to $s$-channel exchange of CP-odd neutral Higgs bosons, leading to constraints on parts of parameter space where the boson in question is light and its couplings to down-type fermions are enhanced, such as occurs in the Type-II two-Higgs-doublet model (2HDM)~\cite{Branco:2011iw}.

In this paper we consider the decay of scalar bottomonium $\chi_{b0} \to \tau^+\tau^-$, which is sensitive to $s$-channel exchange of CP-even neutral Higgs bosons.  This process was first proposed as a probe of Higgs bosons in Ref.~\cite{Haber:1978jt}.  The advent
of the very high luminosity SuperKEKB $e^+e^-$ collider~\cite{Abe:2010gxa} offers the possibility of using $\chi_{b0}\to \tau^+\tau^-$ to put meaningful constraints on the parameter space of the Type-II 2HDM.  To explore this possibility we first estimate the $\chi_{b0}$ decay constant in Sec.~\ref{sec:f_chi}, 
and use it to calculate the relevant branching ratios in Sec.~\ref{sec:decay}.  
We find that the process $\Upsilon(3S) \to \gamma \chi_{b0}(2P) \to \gamma \tau^+\tau^-$ is the most promising, 
though the event rate from SM Higgs exchange is a few orders of magnitude too small to be observed. 
 In Sec.~\ref{sec:doublets} we consider the prospects in the Type-II 2HDM, 
 in which the scalar couplings to both the $b\bar b$ initial state and the $\tau^+\tau^-$ final state 
 can be enhanced.  We show that, with 250~fb$^{-1}$ of data on the $\Upsilon(3S)$, this process has 
 the potential to constrain a large region of as-yet-unexcluded parameter space in which the second 
 CP-even Higgs boson is lighter than about 80~GeV.  
This amount of integrated luminosity represents only a small fraction of the expected ultimate SuperKEKB integrated luminosity of 50~ab$^{-1}$~\cite{Abe:2010gxa}, and could be collected during early running as the accelerator luminosity is ramped up.
 We briefly conclude in Sec.~\ref{sec:summary}.

%%%%%%%%%%%%%%%%%%%%%%%%%%%%%%%%%%%%%%%%%%%%%%
\section{Calculation of the $\chi_{b0}$ Decay Constant}
\label{sec:f_chi}

We use the quark model to calculate the $\chi_{b0}$ decay constant ($f_{\chi_0}$), 
in particular the {\it mock meson} approach
\cite{Hayne:1981zy,Capstick:1989ra,Godfrey:1985pa,Blundell:1995au} 
which has been a useful tool in calculating hadronic matrix elements (see also Ref.~\cite{VanRoyen:1967nq}).  The basic premise
of the quark model is that hadrons are made of constituent quarks and antiquarks
and one 
solves a Schrodinger (like) equation, typically employing a short distance Lorentz vector 
one-gluon-exchange interaction and a Lorentz scalar confining interaction, to obtain
hadron masses and wavefunctions which are used to calculate hadron properties (see for
example Refs.~\cite{Godfrey:1985xj,Godfrey:2015dia}).  Quark model predictions have been
reasonably reliable in describing the properties of known mesons.  Ideally one would 
like to use ab-initio lattice QCD calculations to calculate hadron properties but we know 
of no lattice calculation of $f_{\chi_0}$ so we turn to the quark model.  We note
that quark model predictions~\cite{Godfrey:2015dia} 
give similar agreement to lattice calculations~\cite{Colquhoun:2014ica} for 
the related $\Upsilon$ leptonic decays.  There are a few predictions of scalar decay
constants using the QCD sum-rules approach~\cite{Colangelo:2002mj,Veliev:2010gb,Azizi:2012mp} 
but it is not clear how to relate their 
conventions to the ones used in this paper making comparisons difficult.

The basic assumption of the mock meson approach is that physical hadronic
amplitudes can be identified with the corresponding quark model amplitudes
in the weak binding limit of the valence quark approximation.  
This correspondence is exact in the limit of zero binding and in
the hadron rest frame.  Away from this limit the amplitudes are not in
general Lorentz invariant by terms of order $p_i^2 /m_i^2$.
In this approach the mock meson, $\widetilde{M}$, is 
defined as a state of a free quark and antiquark with the wave function of the
physical meson, $M$:
\begin{widetext}
\begin{equation}
\vert \widetilde{M}(\vec K ) \rangle = \sqrt{2\widetilde{M}_M }
\int d^3p \; \Phi_M (\vec{p} ) \chi_{s\bar s} \phi_{q\bar q} \,
\vert q[ (m_q/\mu )\vec{K} + \vec{p},s]\; \bar{q} [(m_{\bar q}/\mu) \vec{K}
-\vec{p}, \bar{s} ]\rangle,
\label{eqn:mockmeson}
\end{equation}
\end{widetext}
where $\Phi_M(\vec{p})$, $\chi_{s\bar{s}}$ and $ \phi_{q\bar q}$ are momentum,
spin and flavour wave functions respectively, $\mu=m_q +m_{\bar q}$, 
$\widetilde{M}_M$
is the mock meson mass, and
$\vec{K}$ is the mock meson momentum.
To calculate the hadronic amplitude, the physical
matrix element is expressed in terms of Lorentz covariants with 
Lorentz scalar coefficients $A$.  In the simple cases when the mock-meson 
matrix element has the same form as the physical meson amplitude
we simply take $A$ = $\widetilde A$.  

We write the scalar decay constant for the $\chi_0$ meson  as:
\begin{equation}
\langle 0 \vert \bar{q} q \; \vert
\; M(\vec{K}) \rangle =   if_{\chi_0},
\label{eqn:decconstant}
\end{equation}
where  $M \equiv \chi_0$ and the amplitude has been normalized to one particle per unit volume
\cite{VanRoyen:1967nq}.
To calculate the left-hand side of Eq.~(\ref{eqn:decconstant}) we first calculate
\begin{equation}
\langle 0 \vert \bar{q}  q
\;\vert q[(m_q /\mu) \vec{K} + \vec{p},s]\; \bar{q} [(m_{\bar q}/\mu) \vec{K} 
-\vec{p}, \bar{s} ]\rangle 
\end{equation}
using free quark and antiquark wavefunctions, then weight the resulting
expression with the  meson's momentum space wavefunction using 
Eq.~(\ref{eqn:mockmeson}).

There are typically a number of ambiguities in this approach that we must deal with.  
For example, there are several different prescriptions for the 
definition of the meson mass $\widetilde{M}_M$ appearing in Eq.~(\ref{eqn:mockmeson}).
We will use the physical mass as $\widetilde{M}_M$ was introduced  to
give the correct relativistic normalization of the meson's wavefunction and hence
to give the correct kinematics for the process being studied.  Fortunately, many of these ambiguities do not show up in calculations of scalar decay constants, and for the heavy
$b$-quark the non-relativistic limit is a good approximation.  Discussion of these
ambiguities is given in Refs.~\cite{Capstick:1989ra,Godfrey:1985pa,Blundell:1995au}.

Evaluating Eq.~(\ref{eqn:decconstant}) in the non-relativistic limit we obtain
\begin{equation}
f_{\chi_0}=-\frac{3\sqrt{3 M_{\chi_0}}}{\sqrt{\pi} \widetilde m_q} R^{\prime}(0),
\label{eq:fchi}
\end{equation}
where $R^{\prime}(0)$ is the derivative of the radial part of the $\chi_{0}$ wavefunction at the origin, $M_{\chi_{0}}$ is
the measured $\chi_{0}$ mass, and $\widetilde m_q$ is the heavy quark constituent mass.  

The radial wavefunctions are computed using the relativized quark model~\cite{Godfrey:1985xj}.  
For $\chi_{b0}(1P)$ and $\chi_{b0}(2P)$ these wavefunctions were recently computed 
in Ref.~\cite{Godfrey:2015dia}, which found $R^{\prime}_{\chi_{b0}(1P)}(0)=2.255$~GeV$^{5/2}$ 
and $R^{\prime}_{\chi_{b0}(2P)}(0)=2.290$~GeV$^{5/2}$.  For $\chi_{c0}(1P)$ we compute the wavefunction 
following the same procedure and obtain $R^{\prime}_{\chi_{c0}(1P)}(0)=0.912$~GeV$^{5/2}$.  

Inserting these values into Eq.~(\ref{eq:fchi}) and using constituent quark masses $\widetilde m_b=4.977$~GeV 
and $\widetilde m_c = 1.628$~GeV \cite{Godfrey:1985xj}, we obtain
\begin{eqnarray}
f_{\chi_{c0}(1P)} & = & -3.03\; \hbox{GeV}^2, \nonumber \\
f_{\chi_{b0}(1P)} & = & -4.17\; \hbox{GeV}^2, \nonumber \\
f_{\chi_{b0}(2P)} & = & -4.31\; \hbox{GeV}^2.
\end{eqnarray}

%%%%%%%%%%%%%%%%%%%%%%%%%%%%%%%%%%%%%%%%%%%%%%%%
\section{$\chi_0 \to \ell^+\ell^-$ decay and SM event rates}
\label{sec:decay}

Now that we have an estimate for the scalar quarkonium decay constants we can 
obtain expressions for their decay widths and branching ratios.  The matrix element for a scalar
meson $\chi_0$ decaying to two leptons $\ell^+\ell^-$ via an $s$-channel SM Higgs boson is given by
\begin{eqnarray}
\mathcal{M}^H &=& \langle \ell^+\ell^- | \frac{m_\ell}{v} \bar{\ell}\ell | 0 \rangle
{i\over{M_H^2}} \langle 0 | \frac{m_q}{v} \bar{q}q | \chi_0 \rangle \nonumber \\
&=& - \left(\frac{m_q m_{\ell}}{v^2 M_H^2}\right) \; f_{\chi_0} \bar u(p_{\ell^-}) v(p_{\ell^+}),
\label{eq:matrixelt}
\end{eqnarray}
where $v^2 = 1/\sqrt{2} G_F$ is the SM Higgs vacuum expectation value, $M_H$ is the Higgs mass, we have neglected $M_{\chi_0}^2$ relative to $M_H^2$ in the propagator, and in the second line we have used Eq.~(\ref{eqn:decconstant}).
The partial width is then given by
\begin{equation}
\Gamma^H(\chi_0 \to \ell^+\ell^-) = \frac{M_{\chi_0}}{8\pi} 
	\left[1 - \frac{4m_{\ell}^2}{M_{\chi_0}^2} \right]^{3/2} 
				\left( { {m_q m_\ell}\over {v^2 M_H^2} } \right)^2 f_{\chi_0}^2.
\label{eq:partialwidth}
\end{equation}
This expression is in agreement with those given by Refs.~\cite{Haber:1978jt,Barger:1987xg} 
with the exception that
those papers take the current and constituent quark masses to be equal to each other. 

There is also a contribution to the $\chi_0 \to \ell^+ \ell^-$ decay through a two-photon intermediate state.  Following Ref.~\cite{Rashed:2010jp}, we estimate the partial width for this one-loop process using the optical theorem,
\begin{equation}
	\Gamma^{2\gamma}(\chi_0 \to \ell^+\ell^-) \simeq \frac{\alpha^2}{2 \beta_{\ell}} 
		\left[ \frac{m_{\ell}}{M_{\chi_0}} \ln \frac{(1 + \beta_{\ell})}{(1 - \beta_{\ell})} \right]^2
		\Gamma(\chi_0 \to \gamma\gamma),
\end{equation}
where $\alpha$ is the electromagnetic fine structure constant,
\begin{equation}
	\Gamma(\chi_0 \to \gamma\gamma) = \frac{4 \pi \alpha^2}{81 M_{\chi_0}^3} f_{\chi_0}^2,
	\quad {\rm and} \quad
	\beta_{\ell} = \sqrt{1 - \frac{4 m_{\ell}^2}{M_{\chi_0}^2}}.
\end{equation}

We first consider $\chi_{c0}(1P) \to \mu^+\mu^-$ (the decay to $\tau^+\tau^-$ is kinematically forbidden). Taking $M_{\chi_{c0}(1P)}=3.415$~GeV, $m_c=1.27$~GeV, $m_\mu=0.10566$~GeV,
and $M_H=125$~GeV~\cite{Agashe:2014kda},
we obtain $\Gamma^H(\chi_{c0}(1P) \to \mu^+\mu^- )=2.5 \times 10^{-20}$~GeV.
Combining with the total width $\Gamma_{\chi_{c0}(1P)}^{\rm tot}=10.5 \pm 0.6$~MeV \cite{Agashe:2014kda},
we obtain the branching ratio from SM Higgs exchange,
\begin{equation}
	{\rm BR}^H(\chi_{c0}(1P) \to \mu^+\mu^-) = 2.4 \times 10^{-18},
\end{equation} 
which is clearly too small to be observed due to the small values
of the muon and charm-quark masses.  The competing two-photon 
intermediate state in fact yields a much larger contribution,
\begin{equation}
	{\rm BR}^{2\gamma}(\chi_{c0}(1P) \to \mu^+\mu^-) \simeq 2 \times 10^{-10}.
\end{equation}

The situation is not quite so dire for the $\chi_{b0}$ decays.  Taking 
$M_{\chi_{b0}(1P)}=9.860$~GeV, $M_{\chi_{b0}(2P)}=10.232$~GeV,
$m_b=4.67$~GeV, and $m_\tau=1.77682$~GeV~\cite{Agashe:2014kda}, we obtain
the SM Higgs-exchange contributions,
$\Gamma^H(\chi_{b0}(1P)\to \tau^+\tau^- )=4.3\times 10^{-16}$~GeV
and  
$\Gamma^H(\chi_{b0}(2P)\to \tau^+\tau^- )=4.8\times 10^{-16}$~GeV.

To estimate the branching ratios we will need the total widths
for the $\chi_{b0}(1P)$ and $\chi_{b0}(2P)$, which have not been measured. To estimate them we 
follow Ref.~\cite{Godfrey:2015dia}, which combined the measured branching ratios for
$\chi_{b0} \to \gamma \Upsilon(1S)$ with the predicted partial widths for these 
transitions, yielding $\Gamma_{\chi_{b0}(1P)}^{\rm tot}=1.35$~MeV and 
$\Gamma_{\chi_{b0}(2P)}^{\rm tot}= (247\pm 93)$~keV~\cite{Godfrey:2015dia}.  
Note that there is modelling dependence in these estimates 
as we have relied on the results of the relativized quark model to obtain the radiative transition
widths.  Furthermore, the experimental values for the $\chi_{b0} \to \gamma \Upsilon(1S)$ branching ratios have experimental errors,
which are particularly large for the $\chi_{b0}(2P)$ state.  
Combining our $\chi_{b0} \to \tau^+\tau^-$ partial width calculation with these total width estimates we obtain the branching ratios from SM Higgs exchange,
\begin{eqnarray}
{\rm BR}^H(\chi_{b0}(1P) \to \tau^+\tau^-) &=& 3.1\times 10^{-13}, \nonumber \\ 
{\rm BR}^H(\chi_{b0}(2P) \to \tau^+\tau^-) &=& (1.9 \pm 0.5) \times 10^{-12}.
\label{eq:BRs}
\end{eqnarray}
The competing two-photon intermediate state again yields a larger contribution,
\begin{eqnarray}
{\rm BR}^{2\gamma}(\chi_{b0}(1P) \to \tau^+\tau^-) &\simeq& 1 \times 10^{-9}, \nonumber \\
{\rm BR}^{2\gamma}(\chi_{b0}(2P) \to \tau^+\tau^-) &\simeq& 6 \times 10^{-9}.
\label{eq:2ga}
\end{eqnarray}

The final step is to estimate event rates to see if these processes are actually measurable.  
We base our estimates on the production of the $\Upsilon (2S)$ and
$\Upsilon (3S)$ at the high luminosity SuperKEKB $e^+e^-$ collider followed by a 
radiative decay to the $\chi_{b0}$ states.  The $e^+e^- \to \Upsilon(2S)$ cross section
averaged over the Belle and BaBar measurements is about $6.5$~nb \cite{Bevan:2014iga}.
Assuming $\mathcal{L} = 250$~fb$^{-1}$ of
integrated luminosity yields $1.6 \times 10^{9}$ $\Upsilon(2S)$'s produced.  The branching ratio
for $\Upsilon (2S) \to \gamma \chi_{b0}(1P)$ is $(3.8 \pm 0.4)$\%~\cite{Agashe:2014kda}, 
which would yield $6.2 \times 10^7$ $\chi_{b0}(1P)$'s.  
Combining this with the branching ratios in Eqs.~(\ref{eq:BRs}--\ref{eq:2ga}) 
yields $1.9 \times 10^{-5}$ $\gamma \tau^+\tau^-$ signal events from SM Higgs exchange and
about 0.07 events from the two-photon intermediate state.

Likewise we can estimate the number of $\chi_{b0}$'s
that would be produced from $\Upsilon (3S)$ decay. 
The $\Upsilon(3S)$ $e^+e^-$ production cross section is 4~nb \cite{Bevan:2014iga}, and with  
${\cal L}=250$~fb$^{-1}$  of integrated luminosity this yields $10^9$ $\Upsilon(3S)$'s.  
The branching ratio for $\Upsilon(3S)\to \gamma \chi_{b0}(2P)$ is $(5.9\pm 0.6)$\%
and for $\Upsilon(3S)\to \gamma \chi_{b0}(1P)$ is $(0.27 \pm 0.04)$\%~\cite{Agashe:2014kda},
which yield $5.9\times 10^7$ $\chi_{b0}(2P)$'s and $2.7\times 10^6$ 
$\chi_{b0}(1P)$'s.  Combining these with the branching ratios in Eqs.~(\ref{eq:BRs}--\ref{eq:2ga}) 
yields $1.1 \times 10^{-4}$ and $8.5 \times 10^{-7}$ $\gamma \tau^+\tau^-$ signal 
events via an $s$-channel SM Higgs boson, for the decays via $\chi_{b0}(2P)$ and $\chi_{b0}(1P)$ respectively.  The number of signal events from the two-photon intermediate state is about 0.3 and 0.003 for the decays via $\chi_{b0}(2P)$ and $\chi_{b0}(1P)$ respectively.

Clearly these event numbers are too small to be able
to observe $\chi_{b0}\to \tau^+\tau^-$ decays mediated by the SM Higgs boson.  
However, if the fermion-Higgs couplings were enhanced these decays might become observable.  
We explore this possibility in the following section.

%%%%%%%%%%%%%%%%%%%%%%%%%%%%%%%%%%%%%%%%%%%%%%
\section{Signal rates in the two-Higgs-doublet model}
\label{sec:doublets}

\subsection{Resonant signal}

To explore the possibility of enhanced $\chi_{b0} \to \tau^+\tau^-$ decays, we consider the Type-II 2HDM~\cite{Branco:2011iw}.  In this model the scalar couplings to $b$ quarks and $\tau$ leptons can be simultaneously enhanced for large values of the parameter $\tan\beta$, which is defined as the ratio of vacuum expectation values of the two Higgs doublets.  The model contains two CP-even neutral scalars, which we call $H_{125}$ and $H_{\rm new}$.  We identify $H_{125}$ with the discovered Higgs boson at 125~GeV.  The Higgs-exchange matrix element in Eq.~(\ref{eq:matrixelt}) gets modified by the presence of the second Higgs resonance, yielding
\begin{equation}
	\left( \frac{m_b m_{\tau}}{v^2 M_H^2} \right)^2 
	\rightarrow \left[ \frac{m_b m_{\tau}}{v^2} 
	\left( \frac{\kappa_b^{125} \kappa_{\tau}^{125}}{M_H^2}
	+ \frac{\kappa_b^{\rm new} \kappa_{\tau}^{\rm new}}{M_{\rm new}^2 - M_{\chi_{b0}}^2} \right) \right]^2,
\end{equation}
where $M_{\rm new}$ is the mass of the second scalar $H_{\rm new}$ and the $\kappa$ factors represent the couplings of the two scalars to $b$ quarks or $\tau$ leptons normalized to the corresponding coupling of the SM Higgs boson~\cite{LHCHiggsCrossSectionWorkingGroup:2012nn}.  We have kept the $p^2 = M_{\chi_{b0}}^2$ dependence in the second diagram because we will be interested in low $M_{\rm new}$.

Setting the couplings of the 125~GeV Higgs boson equal to their SM values (i.e., working in the \emph{alignment limit}~\cite{Gunion:2002zf}), the branching ratios in Eq.~(\ref{eq:BRs}) are modified by the multiplicative factor
\begin{equation}
	\left[ 1 + \frac{M_H^2}{M_{\rm new}^2 - M_{\chi_{b0}}^2} \tan^2\beta \right]^2.
\end{equation}
The number of signal events grows with increasing $\tan\beta$ and decreasing $M_{\rm new}$.  For a large enough enhancement, the $H_{\rm new}$-exchange contribution will dominate over the SM two-photon intermediate state process.  As we will see, a detectable signal will require a large number $\gg 1$ of signal events, so that the SM two-photon contribution can be neglected.

\subsection{A continuum signal?}

There is also a continuum signal from $\Upsilon \to \gamma H_{\rm new}^* \to \gamma\tau^+\tau^-$, in which the photon is not monoenergetic (we do not consider $M_{\rm new} \lesssim 10$~GeV, which is excluded by searches for $\Upsilon \to \gamma H_{\rm new}$~\cite{Wilczek:1977zn,Franzini:1987pv}).  This can be computed from the on-shell $\Upsilon \to \gamma H_{\rm new}$ decay width by taking the Higgs off-shell.  In particular, neglecting the SM Higgs contribution we have
\begin{widetext}
\begin{equation}
	\Gamma(\Upsilon \to \gamma H_{\rm new}^* \to \gamma \tau^+\tau^-) 
	= \frac{1}{\pi} \int_{4 m_{\tau}^2}^{M_{\Upsilon}^2} 
	\frac{dQ^2 \, Q \,
		\Gamma(\Upsilon \to \gamma H_{\rm new}^*) \,
		\Gamma(H_{\rm new}^* \to \tau\tau) }
	{(Q^2 - M_{\rm new}^2)^2 + M_{\rm new}^2 \Gamma_{\rm new}^2},
\end{equation}
\end{widetext}
where $\Gamma_{\rm new}$ is the total width of $H_{\rm new}$ and the partial widths in the numerator are computed by setting the $H_{\rm new}$ mass equal to $\tau^+\tau^-$ invariant mass $Q$.  We use~\cite{Wilczek:1977zn}\footnote{We correct a misprint in Ref.~\cite{Wilczek:1977zn} following Refs.~\cite{Haber:1978jt,Barger:1987xg} (see also Refs.~\cite{Polchinski:1984ag,Pantaleone:1984ug,Faldt:1987zu,Doncheski:1988jr}).}
\begin{equation}
	\Gamma(\Upsilon \to \gamma H_{\rm new}^*) = \frac{(m_b \kappa_b^{\rm new})^2}{2 \pi \alpha v^2} \left[1 - \frac{Q^2}{M_{\Upsilon}^2} \right] \, \Gamma(\Upsilon \to \mu^+\mu^-)
\end{equation}
and
\begin{equation}
	\Gamma(H_{\rm new}^* \to \tau^+\tau^-) = \frac{(m_{\tau} \kappa_{\tau}^{\rm new})^2 Q}{8 \pi v^2} \left[ 1 - \frac{4 m_{\tau}^2}{Q^2} \right]^{3/2}.
\end{equation}

By integrating numerically we can compare the number of continuum signal events to the number of resonant signal events through the intermediate $\chi_{b0}$.  Taking $M_{\rm new}$ to be well above the $\Upsilon$ mass, the dependence on $M_{\rm new}$ and the coupling enhancement factors $\kappa_{b,\tau}^{\rm new}$ is the same in the continuum and resonant processes, and so drops out in their ratio.  We find that, on the $\Upsilon(3S)$ the total continuum signal rate is only about 0.5\% of the resonant rates through the $\chi_{b0}(2P)$ and $\chi_{b0}(1P)$, and on the $\Upsilon(2S)$ the total continuum signal rate is only about 3\% of the resonant rate through the $\chi_{b0}(1P)$.  Furthermore, this small continuum production is spread over a photon energy range of about 6~GeV, compared to the resonant photon peaks with widths of order an MeV or less (see next subsection).  We therefore neglect the continuum signal in what follows.

\subsection{Backgrounds}

The resonant signal is a single photon, monoenergetic in the parent $\Upsilon$ rest frame, with the remainder of the collision energy taken up by the $\tau^+\tau^-$ pair.  This must be discriminated from the reducible background $\Upsilon \to \gamma \chi_{b0}$ with $\chi_{b0}$ decaying to anything other than $\tau^+\tau^-$, as well as from the irreducible continuum background $e^+e^- \to \gamma \tau^+ \tau^-$.  We will assume that the $\tau^+\tau^-$ identification purity will be good enough that the reducible backgrounds can be ignored.  We then only have to worry about the irreducible background in a signal window around the characteristic photon energy.  

In Table~\ref{tab:signals} we give the photon energies $E_{\gamma}$ and natural linewidths $\delta E_{\gamma}$ (computed from the $\chi_{b0}$ total decay widths) for the three signal processes $\Upsilon \to \gamma \chi_{b0} \to \gamma \tau^+\tau^-$ evaluated in the $\Upsilon$ center-of-mass frame.  We also give the differential cross section of the continuum $e^+e^- \to \gamma \tau^+ \tau^-$ background at the corresponding photon energy, for running at the appropriate $\Upsilon$ resonance energy.  The latter was evaluated using MadGraph5\_aMC@NLO~\cite{Alwall:2014hca} with a generator-level cut on the photon rapidity of $|\eta_{\gamma}| < 5$ in the center-of-mass frame.

\begin{table}
\begin{tabular}{cccccc}
\hline \hline
Parent & Daughter & $E_{\gamma}$ & $\delta E_{\gamma}$ & $d\sigma_{B}/d E_{\gamma}$ & $N_B$ \\
\hline
$\Upsilon(3S)$ & $\chi_{b0}(2P)$ & 122~MeV & 0.24~MeV & 36~fb/MeV & 4320 \\
$\Upsilon(3S)$ & $\chi_{b0}(1P)$ & 484~MeV & 1.3~MeV & 8.8~fb/MeV & 5720 \\
$\Upsilon(2S)$ & $\chi_{b0}(1P)$ & 163~MeV & 1.3~MeV & 30~fb/MeV & 19500 \\
\hline \hline
\end{tabular}
\caption{Tagging photon energies $E_{\gamma}$ in the $\Upsilon$ center-of-mass frame for the three processes considered, the linewidth $\delta E_{\gamma}$ of the photon peak, the differential cross section $d\sigma_B/dE_{\gamma}$ of the continuum $e^+e^- \to \gamma \tau^+ \tau^-$ background evaluated at the photon peak, and the number $N_B$ of continuum $e^+e^- \to \gamma \tau^+\tau^-$ events in a window of width $2 \delta E_{\gamma}$ centered at the photon peak in 250~fb$^{-1}$ of integrated luminosity.}
\label{tab:signals}
\end{table}

\subsection{Sensitivity}

To make a conservative first estimate of the sensitivity, we take as background the total number of $e^+e^- \to \gamma\tau^+\tau^-$ events with a photon energy within a window of width $2 \delta E_{\gamma}$.\footnote{This choice of the photon energy window provides good signal efficiency while maintaining near-optimal signal significance in the presence of large backgrounds~\cite{Landsberg:2000ht}.}  The resulting number of background events in the signal window is shown in the last column of Table~\ref{tab:signals}.  We do not include the $\tau$ reconstruction and identification efficiencies.
A more sophisticated event selection based on better modelling of the background, for example taking into account the angular distributions and $\tau$ polarizations, would improve the sensitivity.   

\begin{figure}
\resizebox{0.5\textwidth}{!}{\includegraphics{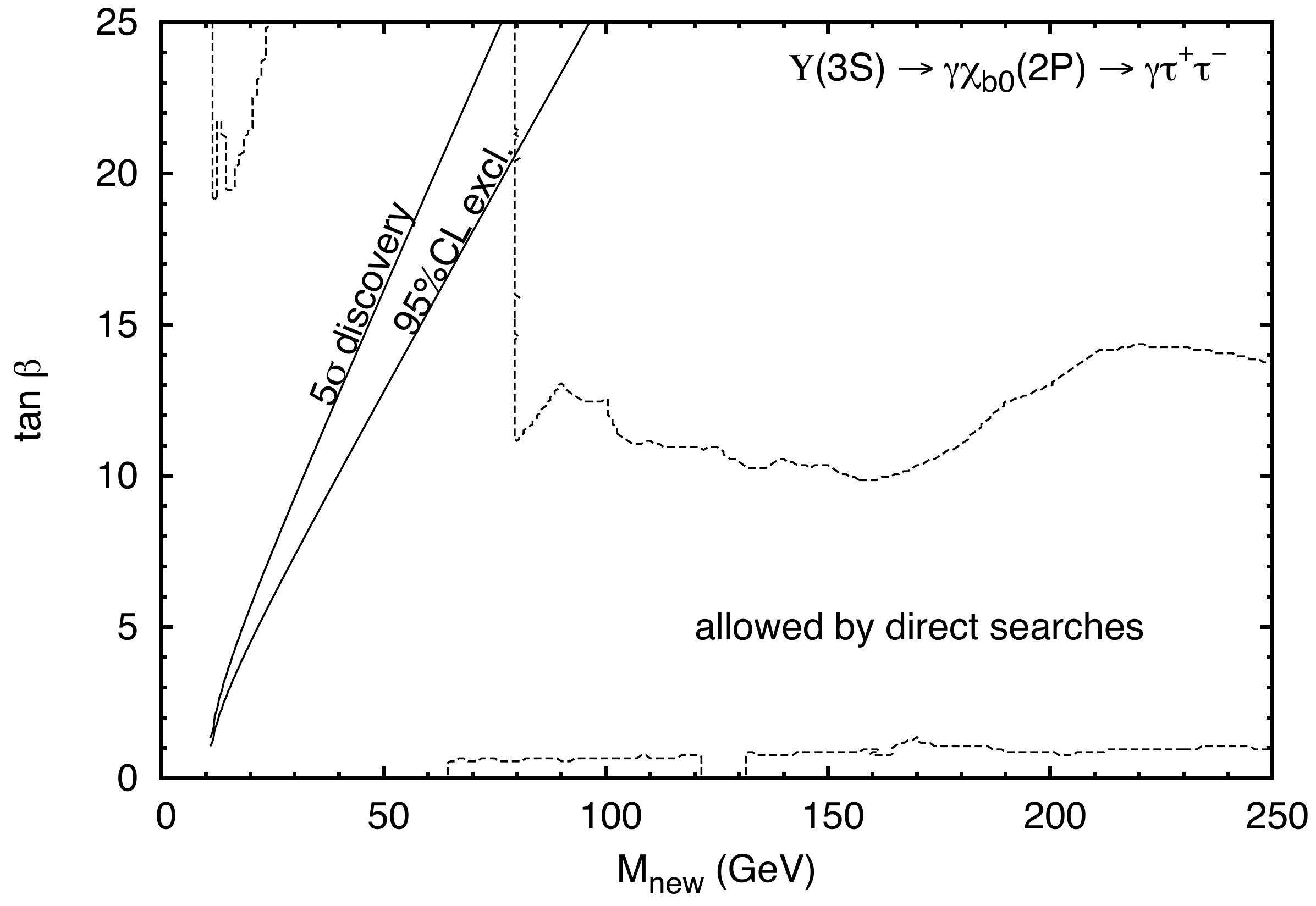}}
\caption{$5\sigma$ discovery and 95\% confidence level (CL) exclusion reach in the Type-II 2HDM from 250~fb$^{-1}$ of data on the $\Upsilon (3S)$.  The sensitivity is to the regions to the left of the solid curves.  We have set the couplings of the 125~GeV Higgs boson equal to their SM values.  The dashed lines indicate the parameter regions still allowed by direct searches for $H_{\rm new}$, computed using HiggsBounds~4.2.0~\cite{Bechtle:2008jh}.}
\label{fig:2P}
\end{figure}

\begin{figure}
\resizebox{0.5\textwidth}{!}{\includegraphics{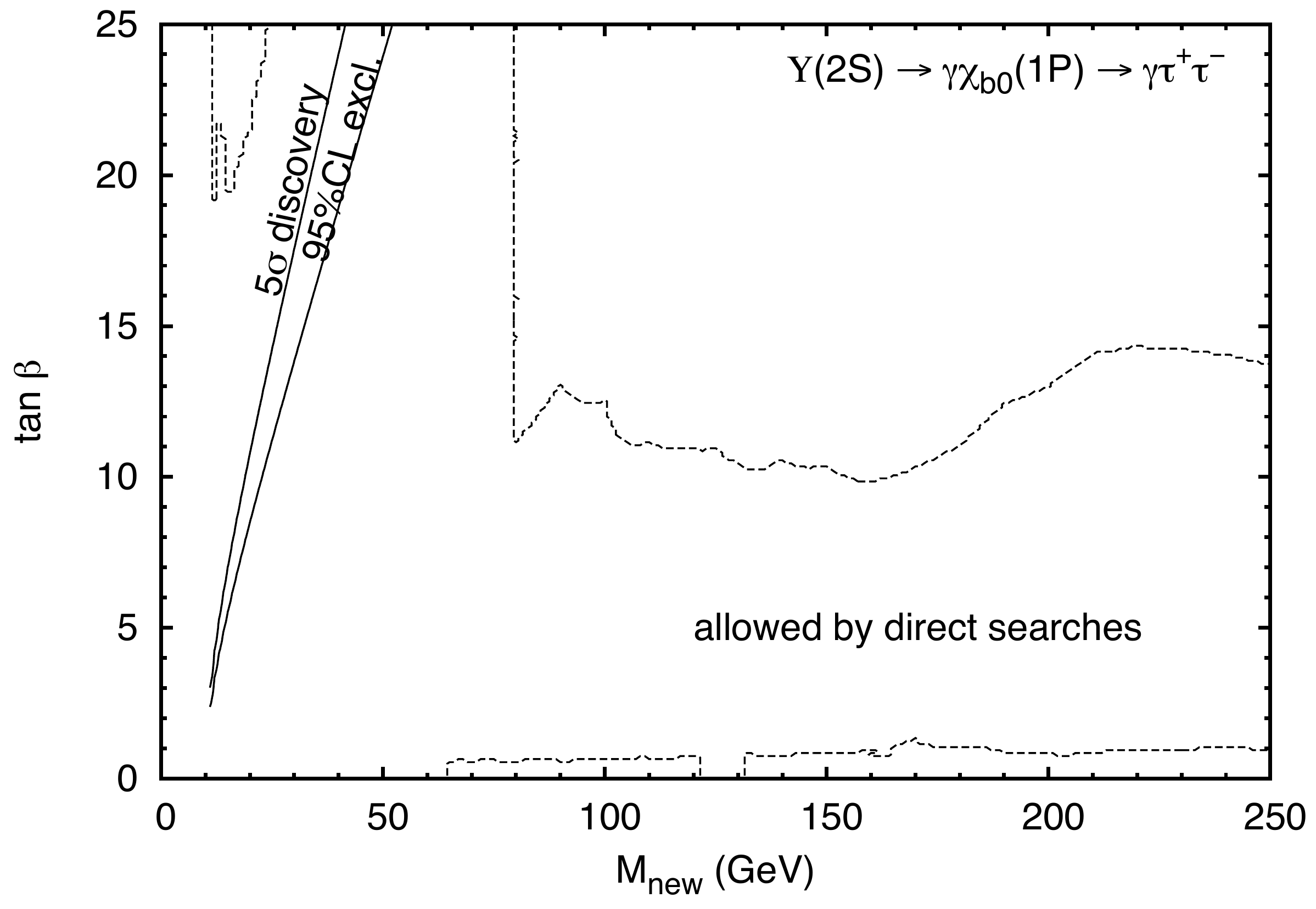}}
\caption{As in Fig.~\ref{fig:2P} but for 250~fb$^{-1}$ of data on the $\Upsilon (2S)$.}
\label{fig:1P}
\end{figure}

In Figs.~\ref{fig:2P} and \ref{fig:1P} we show the resulting $5\sigma$ discovery reach and 95\% confidence level (CL) exclusion reach from 250~fb$^{-1}$ of data on the $\Upsilon(3S)$ and $\Upsilon(2S)$, respectively.  We plot the sensitivity reach as a function of $M_{\rm new}$ and $\tan\beta$, assuming that the couplings of the 125~GeV Higgs boson take their SM values.\footnote{We will further assume that the branching ratio for $H_{125} \to H_{\rm new} H_{\rm new}$ remains small in order to avoid constraints from modifications of the 125~GeV Higgs boson signal strengths.}  We also show, using dotted lines, the parameter region allowed by direct searches, as computed using HiggsBounds~4.2.0~\cite{Bechtle:2008jh}.\footnote{In Figs.~\ref{fig:2P} and \ref{fig:1P} the large excluded region at large $\tan\beta$ and $M_{\rm new} > 80$~GeV is from a CMS $pp \to \phi \to \tau\tau$ search using 7 and 8~TeV data~\cite{Khachatryan:2014wca}.  The small excluded region at large $\tan\beta$ and $M_{\rm new} \sim 10$--25~GeV is from a DELPHI $e^+e^- \to b \bar b \phi \to b \bar b b \bar b$ search for a CP-even $\phi$~\cite{Abdallah:2004wy}.  The exclusion at very low $\tan\beta$ is from an ATLAS scalar diphoton resonance search using 8~TeV data~\cite{Aad:2014ioa}.}  $H_{\rm new}$ masses below about 10~GeV are generally excluded by searches for $\Upsilon \to \gamma H_{\rm new}$~\cite{Wilczek:1977zn,Franzini:1987pv}, which have not been included in HiggsBounds.  We note that the 95\% CL exclusion line for the $\Upsilon(3S)$-initiated process corresponds to 130 signal events on top of about 4300 background events, so that more sophisticated kinematic cuts could improve signal to background substantially and even more so for the $5\sigma$ discovery curves.

On the $\Upsilon(3S)$, the sensitivity comes almost entirely from decays to $\chi_{b0}(2P)$; the signal rate from $\chi_{b0}(1P)$ is more than one hundred times smaller but with comparable background.  This process has the potential to probe a large region of the Type-II 2HDM parameter space with $H_{\rm new}$ masses below 80~GeV and moderate to large $\tan\beta$ that is currently unconstrained by existing searches.  On the $\Upsilon(2S)$ the sensitivity is not as good due to a combination of lower signal rate and a larger photon linewidth, resulting in more background.  Nevertheless, this process can still probe a significant unexcluded region of Type-II 2HDM parameter space for $H_{\rm new}$ masses up to 50~GeV at $\tan\beta = 25$.

%%%%%%%%%%%%%%%%%%%%%%%%%%%%%%%%%%%%%%%%%%%%%%
\section{Summary}
\label{sec:summary}

We have computed the scalar decay constants for $\chi_{c0}(1P)$, $\chi_{b0}(1P)$, and $\chi_{b0}(2P)$, which allowed us to 
predict the branching ratios for decays of these mesons into $\mu^+\mu^-$ or $\tau^+\tau^-$ via an $s$-channel Higgs boson.
While the expected numbers of events from SM Higgs exchange are orders of magnitude too small to observe, 
the $\chi_{b0} \to \tau^+\tau^-$ branching ratio can be significantly enhanced in the Type-II 2HDM.  This leads to 
the first, to our knowledge, indirect probe of the second neutral CP-even scalar from scalar meson decays.

The most promising channel is $\Upsilon(3S) \to \gamma \chi_{b0}(2P) \to \gamma \tau^+\tau^-$.  With 250~fb$^{-1}$ of data collected on the $\Upsilon(3S)$ at SuperKEKB, this process has the potential to constrain a large region of yet-unexcluded Type-II 2HDM parameter space in which the second CP-even neutral Higgs boson is lighter than about 80~GeV.  A more sophisticated rejection of the continuum background should improve this reach further.  Running instead on the $\Upsilon(2S)$ yields fewer signal events and larger backgrounds, but still allows a large region of 2HDM parameter space to be probed.
We hope that this analysis provides further physics motivation for running at energies other than the $\Upsilon(4S)$ during the early stages of SuperKEKB data-taking.

%%%%%%%%%%%%%%%%%%%%%%%%%%%%%%%%%%%%%%%%%%%%%%
\begin{acknowledgments}
This work was supported by the Natural Sciences and Engineering Research Council of Canada.  \end{acknowledgments}
%%%%%%%%%%%%%%%%%%%%%%%%%%%%%%%%%%%%%%%%%%%%%%

%%%%%%%%%%%%%%%%%%%%%%%%%%%%%%%%%%%%%%%%%%%%%

\end{document}